\documentclass[12pt]{article}
\usepackage{graphicx}
\begin{document}
\title{
Transient acceleration 
from a hybrid Chaplygin gas
 }
\author{
Neven Bili\'c$^1$,
Gary B.\ Tupper$^2$
and
Raoul D.~Viollier$^2$
 \\
 $^1$Rudjer Bo\v{s}kovi\'{c} Institute, 10002 Zagreb, Croatia \\
E-mail: bilic@thphys.irb.hr \\
$^2$Institute of 
Department of Physics,
University of Cape Town, \\ Rondebosch 7701, South Africa\\
E-mail: gtupper@science.uct.ac.za, viollier@physci.uct.ac.za
}
\maketitle
%
%\date{\today}
%----------------------------------------------------------------------%
%-----------------------------------------------------------------------
\begin{abstract}
We extend the Chaplygin gas model for dark matter and
dark energy unification by promoting the
Chaplygin gas parameter $A$ to the potential for an extra
scalar with canonical kinetic energy. 
The hybrid model
allows for accelerated Hubble expansion to be a transient
effect around redshift zero. 
\end{abstract}
%
%-----------------------------------------------------------------------
%
%\setlength{\baselineskip}{2\baselineskip}

\section{Introduction}
%\label{intro}
The discovery of accelerated Hubble expansion
\cite{perl1,ries1}
has presented physics with a profound challenge : to account
for a critical universe in which 74\% is in the form of
`dark energy' with an equation of state
$w_{\rm DE}\equiv p_{\rm DE}/\rho_{\rm DE} \simeq - 1$.
Aside from the obvious cosmological constant $\Lambda$, a variety of
models based on `quintessence'
\cite{wett2}
and `k-essence'
\cite{arme3}
have been put forward; for reviews, see
\cite{peeb4}.
One characteristic of the afore-mentioned models is the
assumption that dark energy is separate and distinct from
dark matter which comprises 22\% of the universe with
$w_{\rm DM} = 0$, the remaining 4\% being ordinary baryonic
matter.

A rather different idea, which has been dubbed {\em quartessence}
\cite{mahl5}, is that both dark energy and dark matter are
aspects of a unified theory. The first definite quartessence
model \cite{kamen6,bil7,fab1} was based upon the Chaplygin gas,
an exotic fluid obeying $p = - A/\rho$, which is equivalent
\cite{bil7} to the Dirac-Born-Infeld theory for a D-brane:
\begin{equation}
{\cal{L}}_{\rm DBI} = - \sqrt{A}
\sqrt{1 - g^{\mu \nu}  \theta_{, \mu} \theta_{, \nu} } \, ,
\label{eq001}
\end{equation}
where $\sqrt{A} \sim \Lambda/G$ is the brane tension.
This Lagrangian
yields a perfect fluid stress-energy tensor with the four-velocity
$u_{\mu} =$  $\theta_{,\mu} /
\sqrt{g^{\alpha \beta}  \theta_{, \alpha}  \theta_{, \beta} }$,
the pressure $p = {\cal{L}}_{\rm DBI}$ and the density
$\rho = \sqrt{A} / \sqrt{1 - g^{\mu \nu}
 \theta_{, \mu}  \theta_{, \nu} }$.
Solving the $\theta$ field equation in a Friedmann-Robertson-Walker (FRW)
universe gives
\cite{kamen6}
$\bar{\rho} = \sqrt{ A + B/a^{6} }$, where $B$ is an integration constant
and $a$ the scale factor normalized to unity today. Thus one sees that the
model interpolates between dust at small $a$ and a cosmological constant
at large $a$.

In \cite{bil7} it was suggested that a dominant proportion of the Chaplygin gas condensed on
caustics of the primordial velocity field to provide dark matter haloes,
the residual fraction providing dark energy.
The crucial question is what fraction of the Chaplygin gas goes into condensate.
Indeed, in \cite{bil36} we have noted that if this were 92\% as given by the
geometric Zel'dovich approximation, the CMB and the mass power spectrum
would be reproduced.
However, our analysis of the nonlinear evolution of the Chaplygin gas
\cite{bil8}
demonstrates that  
 the collapse fraction is less than 1\% ,
far too small to significantly affect the mass power spectrum.
This result, 
in qualitative agreement with
that of Avelino et al
\cite{avel35},
is not surprising.
Although the adiabatic speed of
sound $c_{s}^{2} = d p/ d \rho = A/ \rho^{2}$ is small at large redshifts, the comoving
acoustic horizon
$d_{s} = \displaystyle{\int}  dt   c_{s} / a \simeq H_{0}^{-1}   a^{7/2}$ already
reaches Mpc scales at a redshift  $z = 20$, so frustrating condensation.
In the absence of dominant condensation, the Chaplygin gas
model is in gross conflict with the cosmic microwave background
\cite{cart9} and the mass power spectrum
\cite{sand10}. We note that the generalized Chaplygin gas model
\cite{bent11}, $p = - A/\rho^{\alpha}$, with $\alpha < 1$ for
causality, is even less satisfactory because
$d_{s} \simeq \alpha^{1/2} H_{0}^{-1}   a^{2 + 3 \alpha/2}$ is larger.

It has been noted by Reis et al
\cite{reis12} that the root of the structure formation problem is the
term $\Delta \delta p$ in perturbation equations, equal to
$c_{s}^{2} \Delta (\delta \rho / \bar{\rho} )$ for adiabatic
perturbations, and if there are entropy perturbations such that
$\delta p = 0$, no difficulty arises.
This scenario is difficult to justify in the simple  Chaplygin gas model.
Here, we would like to explore a nonadiabatic scenario in which the
constant $A$ is promoted to a field.
In this case,
$\delta p = - \bar{p}   ( \delta \rho / \bar{\rho} - \delta A / \bar{A} )$
and the entropy
perturbations needed to make $\delta p$ vanish
 are attributed to the perturbations of $A$.
  However,  we demonstrate that 
if the kinetic term corresponding to $A$ is standard (or can be made so by a 
field redefinition) the entropy perturbation is automatically 
destroyed on subcausal horizon scales.
 A nonadiabatic scenario with spatial fluctuations of $A$ has recently been
 discussed\footnote{In fact, gauge-invariant perturbations of the 
 single-component model suggested in \cite{zim}  are adiabatic in that
 $\delta p_{\rm GI} = (\bar{A} / \bar{\rho}^{2}) \delta \rho$.
 Moreover, if effectively $A \propto \theta^{- 2n}$ with $0 \leq n < 2$
 to produce acceleration \cite{abra25}, then $d_{s} \simeq H_{0}^{-1} \; 
 a^{2 + 3 (1 - n)/2}$
 showing that such tachyon models share the clustering behaviour of
 the generalized Chaplygin gas.}  
 \cite{zim}
 in the context of the integrated Sachs Wolfe effect.

Irrespective of its other problems, the Chaplygin gas quartessence shares
the feature of eternal acceleration
with most other models of dark energy.
%It is widely believed that string theory should provide a framework for a
% coherent description of our universe.
% However, string theory is incompatible with de Sitter space
% \cite{ban}
The models with eternal acceleration such as $\Lambda$CDM and 
QuintessenceCDM possess a future horizon which  forbids the construction of a consistent S-matrix based on free asymptotic states \cite{ban,hel}.
In this paper we  show that a simple extension of the Chaplygin gas model
yields an end to the accelerating phase and makes this model compatible with
the S-matrix description of string theory.

\section{Hybrid model}
The Lagrangian density for the hybrid Chaplygin gas is
\begin{equation}
{\cal{L}}_{\rm EQ} = \frac{\omega}{2 K}   g^{\mu \nu}   \phi_{, \mu}
\phi_{, \nu} - \sqrt{A}   {\rm e}^{- \phi}
\sqrt{ 1 - g^{\mu \nu}   \theta_{, \mu}   \theta_{, \nu} }\, ,
\label{eq002}
\end{equation}
where
$\omega$ is a coupling constant and the constant
$K = 8 \pi G$
is introduced
to make $\phi$ dimensionless.
%%%%%%%%%%%%%%%%%%%%%%%%%%%%%%%%%%%%%%%%%%%%%%%%%%%%%%%%%%%%%%
That is to say, we have replaced the constant brane
tension,
%%%ADDED BY GBT 11 MARCH
or the Chaplygin gas constant,
%%%%%%%%%%%%%%%%%%%%%%%%%%%%
by the potential for a quintessence-type field $\phi$,
here taken to be exponential.
However,
it is important to stress that,
unlike in quintessence models,
 the field
$\phi$ affects both dark energy and dark matter.
Hence, dark matter and dark energy remain unified as in the
simple Chaplygin gas model.

 Solving the $\theta$ field equation in a
flat FRW model
yields the universe evolution equations in the standard form
\begin{equation}
 H^2\equiv \left(\frac{\dot{a}}{a}\right)^2 =
 \frac{K}{3}(\bar{\rho}_{\theta}+ \bar{\rho}_{\phi})\, ,
\label{eq106}
\end{equation}
\begin{equation}
 \frac{\ddot{a}}{a} =
- \frac{K}{6}(\bar{\rho}_{\theta}+
3\bar{p}_{\theta} + \bar{\rho}_{\phi}  +3 \bar{p}_{\phi})
)\, ,
\label{eq206}
\end{equation}
with
\begin{equation}
\bar{\rho}_{\theta} = \sqrt{A   {\rm e}^{- 2 \phi} + \frac{B}{a^{6}}}\, ,
\label{eq003}
\end{equation}
\begin{equation}
\bar{p}_{\theta} = - \frac{A   {\rm e}^{- 2 \phi}}{\bar{\rho}_{\theta}}\, ,
\label{eq004}
\end{equation}
for the Chaplygin-like component  and
\begin{equation}
\bar{\rho}_{\phi} = \bar{p}_{\phi} = \frac{\omega}{2 K}   \dot{\phi}^{2}
\label{eq005}
\end{equation}
for the $\phi$ field component. In addition, the $\phi$ field satisfies
the equation of motion
\begin{equation}
\ddot{\phi} + 3 H \dot{\phi} = - \frac{K}{\omega}   \bar{p}_{\theta}\, .
\label{eq006}
\end{equation}

\section{Results and discussion}
To preserve the dustlike behaviour of the Chaplygin gas at small $a$,
we require
$\dot{\phi}$ to be small in the small $a$ regime.
In this regime the $\phi$ field equation
(\ref{eq006})
becomes
\begin{equation}
3 H \dot{\phi} \simeq - \frac{K}{\omega}   \bar{p}_{\theta},
\label{eq007}
\end{equation}
whence
%%%EQUATION 10 changed by GBT 11 March
\begin{equation}
\frac{ \bar{\rho}_{\phi}}{\bar{\rho}_{\theta}} \simeq \frac{1}{6 \omega}
\left( \frac{\bar{p}_{\theta}}{\bar{\rho}_{\theta}} \right)^{2} \, .
\label{eq008}
\end{equation}
Solving (\ref{eq007}) with
$\bar{\rho}_{\theta} \simeq \sqrt{B} / a^{3}$ and the initial
conditions\footnote{An initial $\dot{\phi}\neq 0$ will quickly redshift to zero.}
$\phi = \dot{\phi} = 0$ yields
\begin{equation}
\phi (a) \simeq \frac{1}{2}   \ln   \left( 1 + \frac{A}{3\omega B}   a^6 \right) .
\label{eq009}
\end{equation}
Thus, indeed the scalar $\phi$ will be unimportant until
$a \sim 1$ and $| p_{\theta} / \rho_{\theta} | \sim 1$.

\begin{figure}
\begin{center}
\includegraphics[width=.5\textwidth,trim= 0 2cm 0 2cm]{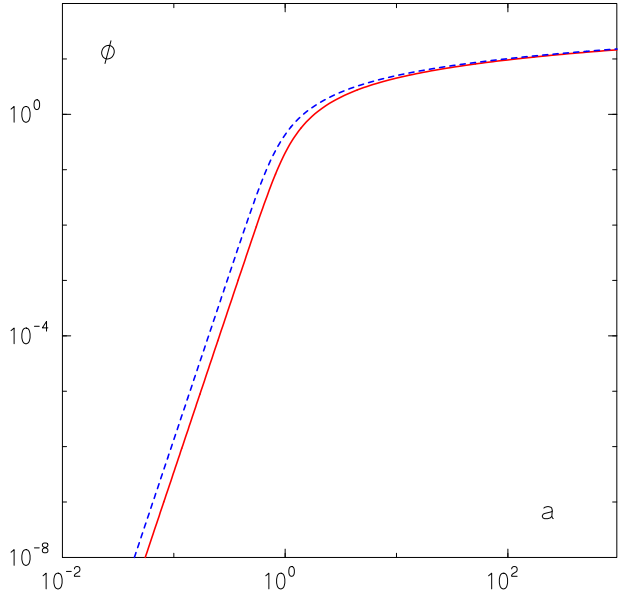}
\caption{
 Evolution of the scalar field $\phi$ for
 $\omega=0.45$ and  the parameter $A/\rho_{\rm cr}^2$ equal to
 0.7 (solid) and 0.9 (dashed line).
}
\label{fig1}
\end{center}
\end{figure}

As in the simple Chaplygin gas, once accelerated expansion starts near $a = 1$,
the quantity
$\dot{\theta}$ is quickly redshifted to zero. For large $a$,
$- \bar{p}_{\theta} \simeq \bar{\rho}_{\theta} \simeq \sqrt{A}   {\rm e}^{- \phi}$,
so the hybrid Chaplygin gas becomes simple quintessence with an exponential
potential. In this large $a$ regime there are simple power law solutions
\begin{equation}
{\rm e}^{- \phi} = \phi_{1}   \left( \frac{t_{1}}{t} \right)^{2};
\;\;\;\;\;
    a = a_{1}   \left( \frac{t}{t_{1}} \right)^{2 \omega},
\label{eq010}
\end{equation}
with $\phi_{1}$, $t_{1}$, and $a_{1}$ being constants,
showing that if $\omega < 1/2$,  deceleration is obtained, and if $\omega = 1/3$,
the behaviour is that of dust.
\begin{figure}
\begin{center}
\includegraphics[width=.5\textwidth,trim= 0 2cm 0 2cm]{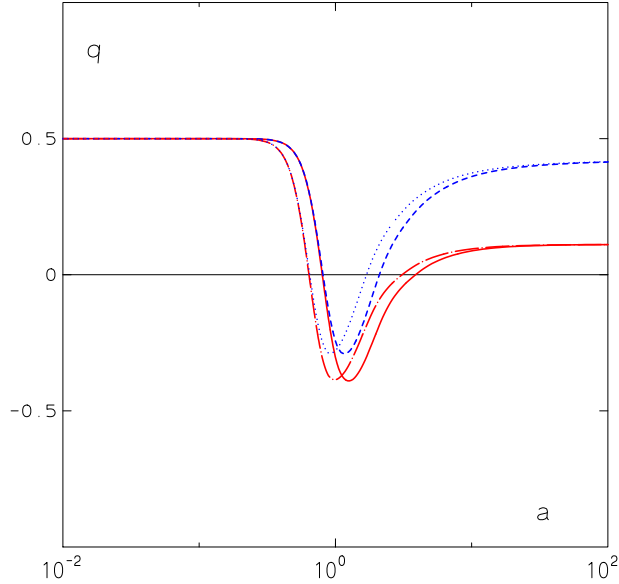}
\caption{
 Deceleration parameter versus $a$ for
  $\omega$ and  $A/\rho_{\rm cr}^2$ equal to,
 respectively,
 0.45 and 0.7 (solid), 0.45 and 0.9 (dot-dashed),
 0.35 and 0.7 (dashed), and 0.35 and 0.9 (dotted line)
}
\label{fig2}
\end{center}
\end{figure}

Hence, the hybrid model with $\omega < 1/2$ supports only a transient
period of accelerated expansion around $a = 1$.
We have confirmed this by numerical calculation: in Fig. \ref{fig1} we exhibit the
solution of Eq. (\ref{eq006}), which fits  Eqs. (\ref{eq009})
 and (\ref{eq010}) well for small and large $a$,
respectively. 
In Fig. \ref{fig2} we show the deceleration parameter
$q = - a \ddot{a} / \dot{a}^{2}$ which briefly goes negative around $a = 1$.

Our hybrid model is in spirit similar to the hybrid qintessence of
 Halyo \cite{hal} since
the field $\phi$ serves as a trigger field responsible for ending acceleration.
As we have already mentioned, this type 
 of behaviour is preferable from the string theory perspective.
A similar transient acceleration has been obtained 
in braneworld models of dark energy
\cite{sah}, in
the two-field supersymmetric hybrid
model \cite{axe}, in double quintessence \cite{bla},  in a
tachyon scenario \cite{cop} and in a variant of the generalized Chaplygin gas
\cite{sen}.

\begin{figure}
\begin{center}
\includegraphics[width=.5\textwidth,trim= 0 2cm 0 3cm]{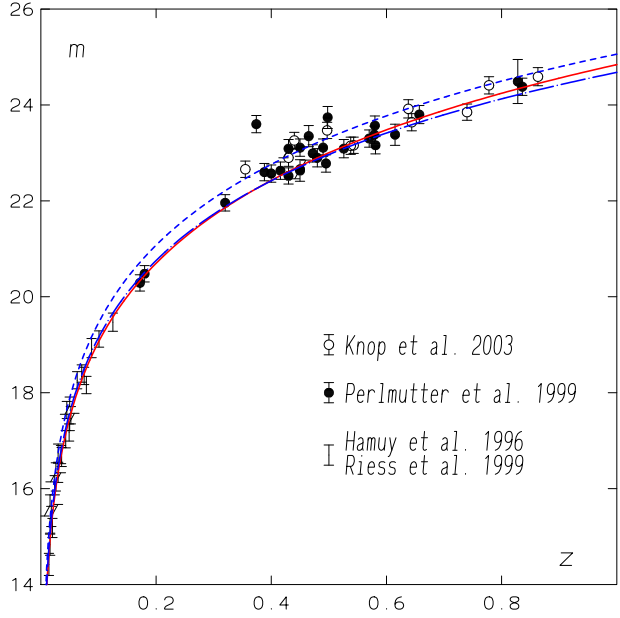}
\caption{
 Magnitude-redshift plot in the
  $\Lambda$CDM model with $\Omega_{\Lambda}=0.7$ and
  $\Omega_{\rm M}=0.3$ (solid line) and in the hybrid
  Chaplygin gas model with $\omega=0.45$ and
$A/\rho_{\rm cr}^2$ equal to
 0.7 (dashed) and 0.9 (dot-dashed line)
 compared with the data taken from the indicated
 references.
 }
\label{fig3}
\end{center}
\end{figure}

For the sake of comparison with the concordance model,
we also calculate the magnitude-redshift relation using the definition
\cite{perl1}
\begin{equation}
m= 5 \log_{10}(H_0 d_{\rm lum}) +M ,
\label{eq108}
\end{equation}
where the luminosity distance for the flat space is defined as
\begin{equation}
d_{\rm lum}=(1+z)\int_0^z \frac{dz'}{H(z')}\,  ,
\label{eq109}
\end{equation}
and $M$ is a constant of the order of 25.
In Fig. \ref{fig3} we plot the results of the calculation in the hybrid 
Chaplygin gas
and in the $\Lambda$CDM model, and compare them with the supernovae type I
observation data
\cite{perl1,ham,ries2,knop}.

Despite the above encouraging features, this extended model does not
solve the structure formation problem.
Indeed, the model of Eq. (\ref{eq002}) is constructed so that
\begin{equation}
\delta   p_{\theta} = - \bar{p}_{\theta}   \left(
\frac{\delta \rho_{\theta}}{\bar{\rho}_{\theta}} - 2 \delta \phi \right)
\label{eq011}
\end{equation}
%%%%%%%%%%%%%%%%%%%%CHANGES HERE
and the entropy perturbations of
\cite{reis12} are realized by $\delta \phi =
 \delta \rho_{\theta} /(2 \bar{\rho}_{\theta})$
as an initial condition outside the causal horizon
$d_{c} = \int dt / a \simeq H_{0}^{-1}  a^{1/2}$. Since $\phi$ is small and the potential very flat, the problem is that $\delta \phi$ behaves essentially as a
free field
%%%%%%%%%%%%%%%%%%%%%%%%%%%%%%%%END
\begin{equation}
\delta \ddot{\phi} + 3 H \delta \dot{\phi} -
\frac{1}{a^{2}}  \Delta \delta \phi \simeq 0   .
\label{eq012}
\end{equation}
%%%CHANGES
Then, once the perturbations enter the horizon $d_{c}$ (but are still outside the
acoustic horizon $d_{s}$ for adiabatic perturbations), $\delta \phi$ undergoes rapid damped
%%%%%%%%%%END
oscillations, so that the entropy perturbation is destroyed. This means that
entropy perturbations are not automatically preserved except on long,  i.e.,
superhorizon, wavelengths where the simple Chaplygin gas has no problem anyway.

%%%%%%%%%%%%%%%NEW PARAGRAPH ADDED 8/3/05
What is wanted for structure formation is a field which avoids the last
term in
Eq. (\ref{eq012})\footnote{Here we are speaking of 
a strictly 4-dimensional viewpoint.
It is not excluded that the resolution of the structure formation problem 
lies with the geometric
interpretation of Eq. (\ref{eq001}) 
 as the brane embedding in higher dimensions \cite{bil26}}.
A natural candidate that comes to mind
\cite{bil4}
 is the Kalb-Ramond field which
is intrinsic to string theory and is coupled to the Dirac-Born-Infeld action \cite{pol13}.
Related models have recently been considered in the context
of D-brane world cosmology \cite{chu14}
and phantom energy stars \cite{bil3}.
%%%%%%END OF NEW PARAGRAPH

In conclusion, we have examined a hybrid  model, promoting the
Chaplygin gas constant to the potential for an extra scalar $\phi$. The hybrid
model preserves the main successes, and failures, of Chaplygin quartessence but
has the added feature of future deceleration.

\subsection*{Acknowledgment}
 The work of NB   was supported by
 the Ministry of Science and Technology of the
 Republic of Croatia under Contract
 No. 0098002 and partially supported through the
 Agreement between the Astrophysical sector, SISSA, and the
 Particle Physics and Cosmology Group, RBI.
%%%ADDED BY GBT 11 MARCH
Two of us (RDV and GBT) acknowledge grants from the South
African National Research Foundation (NRF GUN-2053794),
the Research Committee of the University of Cape Town,
and the Foundation for Fundamental Research (FFR PHY-99-01241).
%%%%%%%%%%%%%%%%%%%%%%%%%%%%%%%%%%%%%%%%%%%%%%%%%%%%%%%%%%%%%%%%

\end{document}